\documentclass[prb,twocolumn,superscriptaddress,showpacs,%
preprintnumbers,amsmath,amssymb]{revtex4}
\usepackage{amsmath}
\usepackage{graphicx}
\usepackage{dcolumn}

\usepackage{booktabs}

\begin{document}

\title{Stark shift of excitons and trions in two-dimensional materials}

\author{L. S. R. Cavalcante} \email{lucaskvalcante@fisica.ufc.br}
\author{D. R. da Costa} 
\affiliation{Departamento de F\'isica, Universidade Federal do
Cear\'a, Caixa Postal 6030, Campus do Pici, 60455-900 Fortaleza,
Cear\'a, Brazil}
\author{G. A. Farias}
\affiliation{Departamento de F\'isica, Universidade Federal do
Cear\'a, Caixa Postal 6030, Campus do Pici, 60455-900 Fortaleza,
Cear\'a, Brazil}
\author{D. R. Reichman} 
\affiliation{Department of Chemistry, Columbia University, New York, New York 10027, USA}
\author{A. Chaves} \email{andrey@fisica.ufc.br}
\affiliation{Departamento de F\'isica, Universidade Federal do
Cear\'a, Caixa Postal 6030, Campus do Pici, 60455-900 Fortaleza,
Cear\'a, Brazil}

\date{ \today }

\begin{abstract}
The effect of an external in-plane electric field on neutral and charged exciton states in two-dimensional (2D) materials is theoretically investigated. These states are argued to be strongly bound, so that electron-hole dissociation is not observed up to high electric field intensities. Trions in the anisotropic case of monolayer phosphorene are demonstrated to especially robust under electric fields, so that fields as high as 100 kV/cm yield no significant effect on the trion binding energy or probability density distribution. Polarizabilities of excitons are obtained from the parabolicity of numerically calculated Stark shifts. For trions, a fourth order Stark shift is observed, which enables the experimental verification of hyperpolarizability in 2D materials, as observed in the highly excited states of the Rydberg series of atoms and ions.  
\end{abstract}

\maketitle

\section{Introduction}
A few years after the experimental demonstration of the synthesis and control of mono- and multi-layer graphene \cite{Geim}, a great deal of interest has been focused on a related line of reseach, namely the pursuit of a wider class of stable atomically thin materials, such as transition metal dichalcogenides (TMDCs) \cite{Heinz, ReviewTMDC}, and few-layer black phosphorus (phosphorene) \cite{PaperOriginal,ReviewBP}, among others \cite{Review2Dmaterials}. Much of the interest in these materials lies in their optoelectronic properties, because in contrast with graphene, these materials are naturally semiconductors with highly tunable band structures. One of the most important features ubiquitous to all of these few-layer semiconductors is reduced Coulomb screening, which greatly increases the binding energies of excitons and trions (i.e. charged excitons) \cite{Tony, Ugeda, Thygesen, Berkelbach} when compared to their bulk counterparts. Such strongly bound excitons dominate the photophysical properties and, in principle, enable the investigation of the effect of strong electric fields applied to these systems without intervening electron-hole dissociation. \cite{Footnote} In fact, such tightly bound excitonic complexes in 2D materials have also brought to the fore the possibility of driving a charged exciton with an in-plane field, which would open a new avenue for possible applications in future opto-electronic devices.  

The effect of an applied (static) electric field $F$ on the energy states of hydrogen- and helium-like atoms has been known for decades \cite{Stark}. Such a field modifies the spectrum according to the so-called Stark shift, which for states with $s$-symmetry, is given by
\begin{equation} \label{eq.Stark}
\Delta E = \frac{1}{2}\alpha F^2 + \frac{1}{24}\beta F^4 + \hdots
\end{equation}	
The $\alpha (\beta)$ factor is known as (hyper)polarizability. Since excitons and trions exhibit hydrogen- and helium ion-like electronic states, respectively, one expects to observe such an energy shift in the optical spectrum when an electric field is applied in semiconductor materials as well. The absorption of excitons with $p$-symmetry is not optically allowed by selection rules (except in two-photon experiments \cite{Poem}), therefore, one can use Eq. (\ref{eq.Stark}) for the description of the excitonic Stark effect, where odd-order correction terms are zero due to symmetry. Moreover, corrections with orders higher than $F^4$ are usually negligibly small. Indeed, this hyperpolarizability correction to the Stark shift is usually neglected in the study of excitons under applied fields. Nevertheless, such fourth order corrections to $\Delta E$ have been theoretically investigated by perturbation theory for several atoms and ions, \cite{Bishop} where they are found to be of the order of 10$^{-7}$ cm$^4$/kV$^3$. Hyperpolarizabilities two orders of magnitude higher than this were experimentally observed in highly excited states of the Rydberg series of Ba, Ca and Rb atoms \cite{KulinaBa, KulinaCa, TadaRb}. The tightly bound neutral and charged excitons in 2D materials, where binding energies are much higher than those in previously known cases of excitons in, e.g., heterostructures of III-V and II-VI semiconductors, provides the opportunity to experimentally investigate the (hyper)polarizabilities of e-h complexes.

In this paper, we theoretically investigate the effect of an applied in-plane electric field on the energy states of excitons and trions in 2D materials, namely, TMDCs and $n$-layer black phosphorus ($n$-BP). Energy shifts as a function of the field, as numerically calculated within the effective mass framework and the Wannier-Mott picture, are fitted to the Stark shift expression Eq. (\ref{eq.Stark}). We consider not only the case of suspended 2D materials, but also the situations where the material is deposited on a substrate, or encapsulated by another material, which are the most studied cases. The effect of the dielectric screening from the environment on the exciton and trion Stark shifts in these three situations is discussed. Trions are found to be robust against applied fields. Particularly, trions in $n$-BP barely undergo changes in their binding energies and wave functions if the field is applied along the zigzag direction of the black-phosphorus lattice, which supports the idea that even high applied fields may successfully drive charged excitons across the phosphorene plane with no e-h dissociation. Trion Stark shifts in all cases investigated here turn out to be not properly described by the quadratic Stark shift, as is usual for neutral excitons in these materials, but rather by a quartic shift, where hyperpolarizabilities are found to be even higher than those already experimentally observed in excited states of atoms \cite{KulinaBa, KulinaCa, TadaRb}, as we will demonstrate in further details in what follows.

\section{Theoretical model}
	The exciton Hamiltonian in a system composed by a substrate, a 2D semiconductor layer (where the e-h pair resides), and a capping layer, with dielectric constants $\epsilon_1$, $\epsilon_2$ and $\epsilon_3$, respectively, in the presence of an external electric field $\vec F$, has been given in previous papers \cite{ChavesBP} and is presented here for the sake of completeness:
\begin{equation}\label{eq.excHam}
H_{exc} = \frac{p^2}{2\mu} + V_{eh}(r) + e\vec{F}\cdot\vec{r},
\end{equation}
where $\mu_X^{-1} = m_e^{-1} + m_h^{-1}$, $\vec{r} = \vec{r}_e - \vec{r}_h$ is the e-h relative coordinate, and the interaction potential between particles $i$ and $j$ ($i,j = e$ or $h$) is assumed to be of the Rytova-Keldysh form \cite{Rytova, Keldysh} which takes account of the screening by the environment surrounding the semiconductor layer
\begin{equation}
V_{ij} = \frac{e^2}{4\pi\epsilon_0(\epsilon_1 + \epsilon_3)\rho_0}\left[H_0\left(\frac{\rho}{\rho_0}\right) - Y_0\left(\frac{\rho}{\rho_0}\right)\right]
\end{equation}
with $\rho = |\vec{r}_i - \vec{r}_j|$, $\rho_0 = \epsilon_2d/(\epsilon_1 + \epsilon_3)$, and $d$ the thickness of the semiconductor layer.

\begin{figure}
\centerline{\includegraphics[width=\linewidth]{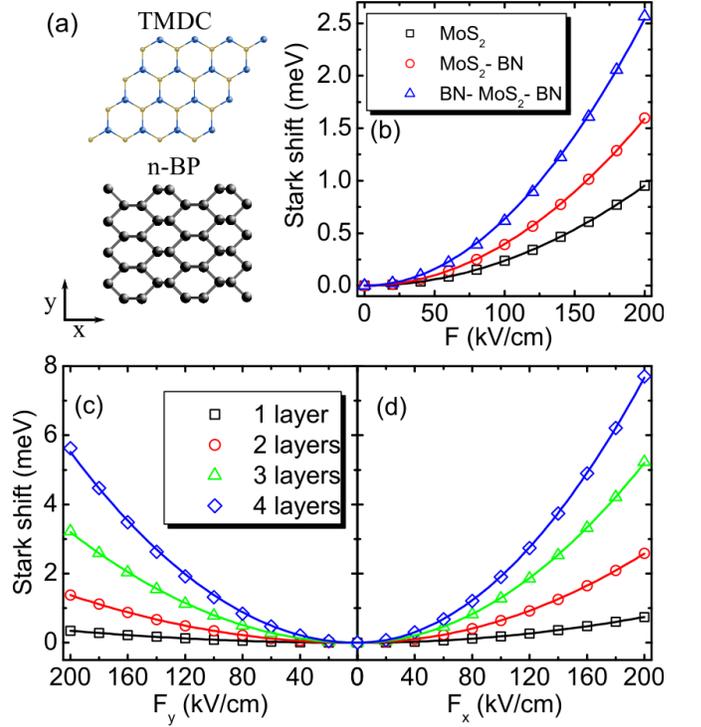}}
\caption{(a) Sketch of the top view of the TMDC and BP crystal lattices, along with the direction of $F_x$ and $F_y$ components of the applied field. Stark shift of exciton binding energies in (b) MoS$_2$, in the suspended, on-substrate, and encapsulated cases, and in (c,d) suspended phosphorene with different numbers of layers. In the latter, results are shown for fields applied in the $y$- and $x$-direction respectively (c,d). The values of polarizability (fitting parameters) are presented in Table \ref{table:TMDCandBP-Exciton}.} 
\label{fig:MoS2-BP-Exciton}
\end{figure}

For trions, we will consider only the negatively charged case, since the results for the positively charged case are similar due to the nearly identical electron and hole effective masses. Such a trion is described as two electrons with mass $m_e$ located at $\vec{r}_e$ and $\vec{r}_{e'}$, and a hole with mass $m_h$ located at $\vec{r}_h$. This six-coordinate system is not easy to implement in numerical calculations and it is thus necessary to find a new coordinates set in which some coordinates are removed by the use of symmetry arguments \cite{ChavesTrions}. Relevant coordinates for this new system are the trion center-of-mass (CM)
\begin{equation}
\vec{R}_t = \frac{m_e \vec{r}_e + m_e \vec{r}_{e'} + m_h \vec{r}_h}{M},
\end{equation}
where  $M = 2m_e + m_h$, the previously defined relative coordinate for an e-h pair forming an exciton $\vec{r}$, and the relative coordinate between the CM of this exciton and the extra electron
\begin{equation}
\vec{u} = \vec{r}_{e'} - \frac{m_e \vec{r}_e + m_h \vec{r}_h}{m_e + m_h}.
\end{equation}
Within the coordinates system $(\vec{R}_t,\vec u, \vec r)$, the trion Hamiltonian reads
\begin{multline}\label{eq.TrionHam}
H_t = \frac{p_r^2}{2\mu_X} + \frac{p_u^2}{2\mu_T} - V_{eh}\left(\vec{r}\right) + V_{ee'}\left(\vec{u} -\gamma_h \vec{r}\right) \\ -V_{e'h}\left(\vec{u} + \gamma_e \vec{r}\right) -  e\vec{F} \cdot \vec{r} - e\vec{F} \cdot \vec{u},
\end{multline}
where $\mu_T^{-1}  = m_e^{-1} + (m_e + m_h)^{-1}$, and $\gamma_e = 1 - \gamma_h = m_e/(m_e + m_h)$. Since the electron-electron and electron-hole interaction potentials, namely $V_{ee'}\left(\vec{u} -\gamma_h \vec{r}\right)$, $V_{eh}\left(\vec{r}\right)$, and $V_{e'h}\left(\vec{u} + \gamma_e \vec{r}\right)$, respectively, do not depend on the trion CM coordinate $\vec{R}_t$, this coordinate is completely separable from the Hamiltonian Eq. (\ref{eq.TrionHam}), so that the Hamiltonian associated to the trion CM motion due to the applied field $H_{CM} = P_{CM}^2/2M + e\vec F \cdot \vec{R}_t$ is left out from $H_t$. Finally, the Schr\"odinger equation for the exciton (trion) with Hamiltonian Eq. (\ref{eq.excHam}) [Eq. (\ref{eq.TrionHam})] is numerically solved by an imaginary time evolution method. \cite{Split-Operator}

\section{Results and discussion}

\begin{figure}
    \centerline{\includegraphics[width=\linewidth]{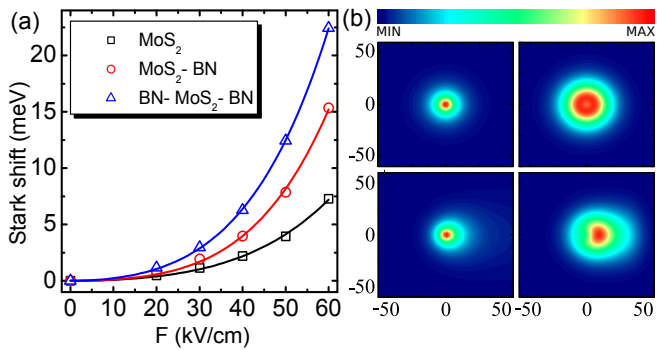}}
    \caption{(a) Numerically obtained (symbols) field dependence of the trion binding energy $E_b$ in monolayer MoS$_2$ in the suspended case, on a BN substrate, and encapsulated by BN. The values of polarizability and hyperpolarizability (fitting parameters for the curves) are presented in Table \ref{table:TMDCandBP-Trion}. (b) Contour map of the square modulus of the trion wave function in suspended MoS$_2$. The left (right) column represents the exciton's (trion's) center of mass wave function in the absence of electric field in the first row, and for a $F = 60$ kV/cm field applied in the $x-$direction in the second row. The colour scale goes from 0 (blue, MIN) to 0.012 (red, MAX).} 
    \label{fig:MoS2-Trion}
\end{figure}

We have calculated the electric field dependence of the exciton binding energies for four different TMDCs, namely MoS$_2$, MoSe$_2$, WS$_2$, and WSe$_2$, as well as for $n$-BP with up to $n$ = 4 layers, for which effective masses and dielectric constants are found in the literature \cite{database, lucas, CastellanosGomez}. Results for TMDCs are independent of the field direction, whereas for the anisotropic case of $n$-BP, we consider fields applied in both $x$ and $y$ directions defined in Fig. \ref{fig:MoS2-BP-Exciton}(a). Substrate and capping layers are assumed to be made of BN. The calculated Stark shift for the case of MoS$_2$ is shown in Fig. \ref{fig:MoS2-BP-Exciton}(b). As the system becomes more screened by the environment, the electron-hole interaction weakens and, as a consequence, the Stark shift in the encapsulated case is more pronounced as compared to the others. As for $n$-BP, a similar screening effect occurs, and it is even more pronounced as the number of layers $n$ increase, as shown for the suspended case in Fig. \ref{fig:MoS2-BP-Exciton}(c,d). The Stark shift in $n$-BP is stronger for fields applied in the $x$-direction direction, where effective masses are lower. Curves in Fig. \ref{fig:MoS2-BP-Exciton}(b-d) are fits of the numerical data (symbols) with the usual expression for Stark shift, which assumes a quadratic dependence on the field, i.e.  Eq. (\ref{eq.Stark}) for $\beta = 0$. Polarizabilities for the four previously mentioned TMDCs and $n$-BP are summarized in Table \ref{table:TMDCandBP-Exciton}, for all cases of dielectric environment. For the suspended case, they agree well with previously reported values \cite{Pedersen, Perebeinos, ChavesBP, Stem}, while results presented here extend this data to the supported (on substrate) and encapsulated cases. In fact, the polarizability for encapsulated WSe$_2$ obtained here is in good agreement with recent experimental results, \cite{Massicote} where it was observed to be $(1 \pm 0.2)\times10^{-6}$ Dm/V, which, converted to the units used here, yields $\approx 10.393 \times 10^{-5}$ meVcm$^2$/kV$^2$, very close to the 10.027$\times 10^{-5}$ meVcm$^2$/kV$^2$ we obtain in our calculations (see Table \ref{table:TMDCandBP-Exciton}).

\begin{figure}
    \centerline{\includegraphics[width=\linewidth]{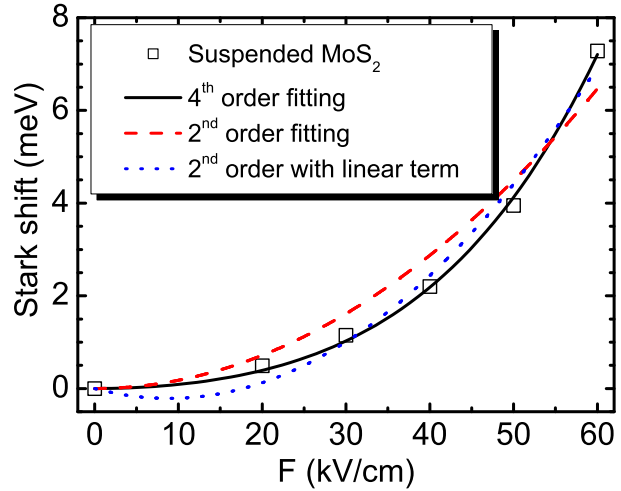}}
    \caption{Comparison between fitting curves of second and fourth order corrections (lines) to the numerically obtained Stark shift of trions (symbols) in suspended MoS$_2$. For the former, two cases are considered, namely with and without a first order term in the fitting expression (see manuscript).} 
    \label{fig:MoS2-Error}
\end{figure}

\begin{table}
\caption{Exciton polarizabilities for monolayer TMDC and $n$-BP in the suspended, on substrate and encapsulated cases, in units of $10^{-5}$ meVcm$^2$/kV$^2$. Results inside (outside) the brackets for $n$-BP are for electric fields applied in $x$ and $y$ directions, respectively.}
{\def\arraystretch{2}\tabcolsep=4pt
\begin{tabular}{c c c c}
\hline\hline
& & Exciton & \\
Material & Suspended & On substrate & Encapsulated \\
\hline
MoS$_2$ & 2.378 & 4.218 & 6.340 \\
MoSe$_2$ & 2.679 & 4.259 & 6.532 \\
WS$_2$ & 2.916 & 6.034 & 9.079 \\
WSe$_2$ & 3.492 & 6.116 & 10.027 \\
1-BP & 1.861 (0.871) & 3.628 (1.575) & 6.162 (2.598)  \\
2-BP & 6.447 (3.440) & 10.045 (5.293) & 14.020 (7.632)  \\
3-BP & 13.005 (7.992) & 17.477 (11.397) & 21.670 (15.684) \\
4-BP & 19.169 (13.786) & 23.467 (18.947) & 27.194 (25.581) \\
\hline\hline
\end{tabular}
}
\label{table:TMDCandBP-Exciton}
\end{table}

\begin{figure}
\centering
\includegraphics[width=\linewidth]{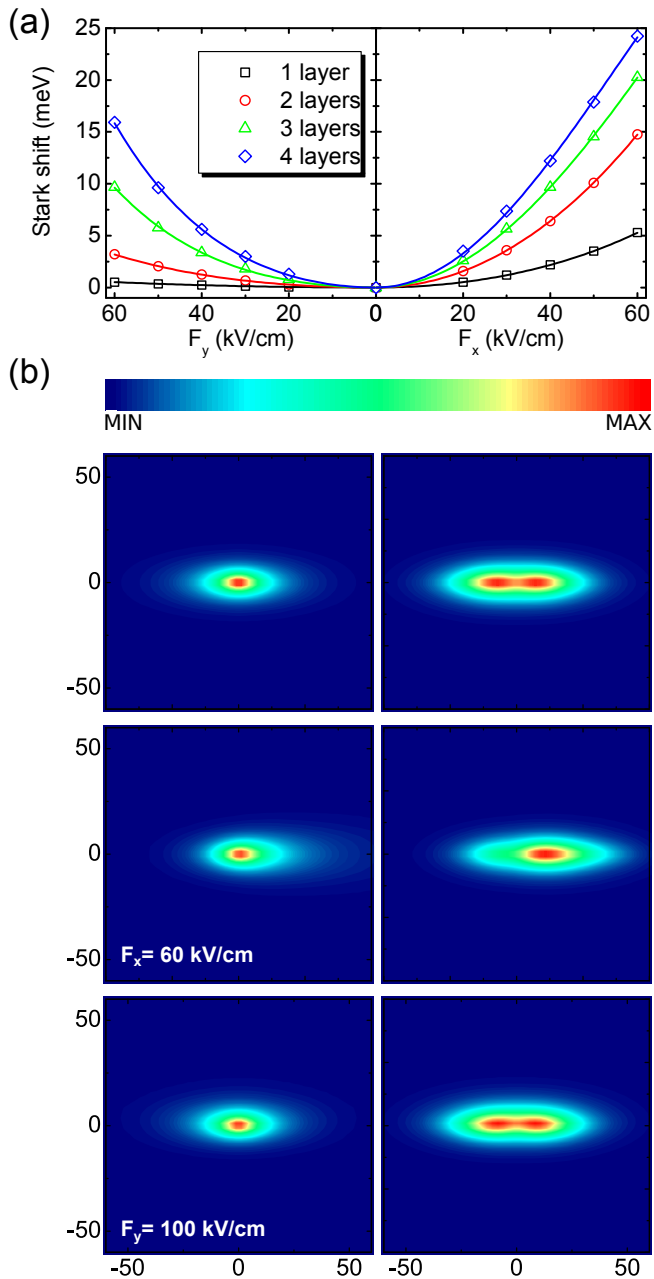}
\caption{(a) Numerically obtained (symbols) electric field dependence of the binding energy $E_b$ for trions in phosphorene with different number of layers for a field applied in the $y$ (left) and $x$ (right) directions. The values of polarizability and hyperpolarizability (fitting parameters for the curves) are presented in Table \ref{table:TMDCandBP-Trion}. (b) Countour maps of the square modulus of the trion wave function in suspended monolayer BP. The left (right) column represents the exciton (trion) center-of-mass wave function in the absence of electric field in the first row, and for a $F = 60$ kV/cm field applied in the $x$-direction in the second row, and $F = 100$ kV/cm applied in the $y$-direction in the third row. The colour scale ranges from 0 (blue, MIN) to 0.018 (red, MAX)} 
\label{fig:BP-trion}
\end{figure}

Trion binding energies in the absence of electric fields for TMDCs and $n$-BP, as obtained by our numerical method, are summarized in Table \ref{table:TMDCandBP-TrionBindingEnergy}. In fact, results for some TMDCs and few layer BP, in the suspended case and supported by a substrate, have been already reported by different papers in the literature. \cite{Tony, Berkelbach, ChavesTrions, VanDerDonck} Our results are in good agreement with these previously reported values and are shown here for completeness. Differences of a few meV are mostly due to the use of slightly different parameters such as effective masses and dielectric constants. Table \ref{table:TMDCandBP-TrionBindingEnergy} also provides results for the case where these materials are encapsulated by BN. The interest in trions in such encapsulated 2D semiconductors has been recently enhanced by experimental reports \cite{Strauf, ZhipengLi} showing that such encapsulation renders trion peaks sharper and clearer in photoluminescence experiments.
\begin{table}
\caption{Trion binding energies (in meV), in the absence of electric fields, for monolayer TMDCs and $n$-BP in the suspended case, on a BN substrate, and encapsulated by BN, respectively.}
{\def\arraystretch{2}\tabcolsep=4pt
\begin{tabular}{c c c c}
\hline\hline
Material & Suspended & On substrate & Encapsulated \\
\hline
MoS$_2$ & 32.12 & 21.59 & 15.83 \\
MoSe$_2$ & 32.00 & 22.60 & 17.18 \\
WS$_2$ & 34.49 & 23.60 & 15.94 \\
WSe$_2$ & 33.15 & 21.93 & 16.14 \\
1-BP & 51.77 & 32.65  & 22.82   \\
2-BP & 27.99  & 19.34  & 14.56   \\
3-BP & 19.18 & 13.97  & 11.10  \\
4-BP & 15.15 & 11.69 & 9.80  \\
\hline\hline
\end{tabular}
}
\label{table:TMDCandBP-TrionBindingEnergy}
\end{table}

The trion Stark shift as a function of the applied field in MoS$_2$ is shown in Fig. \ref{fig:MoS2-Trion}(a), assuming the same three cases of dielectric environments as in the previous discussion. Qualitatively similar results are found for the other TMDCs. It becomes evident that for trions, the assumption of a quadratic shift is no longer valid, as reasonable fittings shown by the curves in Fig. \ref{fig:MoS2-Trion}(a) are only obtained if one assumes a significant contribution of the hyperpolarizability (quartic) term, i.e. $\beta \neq 0$ in Eq. (\ref{eq.Stark}). Moreover, Fig. \ref{fig:MoS2-Trion} illustrates the robustness of the trion state, where no dissociation is observed even for fields up to 60 kV/cm, as shown by the wave function projections in Fig. \ref{fig:MoS2-Trion}(b), for the suspended MoS$_2$ case. Notice that, in the absence of fields (top row of panels), the excitonic and trionic components of the wave function are both circularly symmetric, and the latter exhibits a minimum at $r = 0$, as a consequence of electron-electron repulsion. The 60 kV/cm field induces a deformation of these distributions, as observed in the bottom row of panels, but they remain concentrated around the origin, representing the absence of trion dissociation. 

\begin{table*}
\caption{Trions polarizabilities / hyperpolarizabilities for monolayer TMDCs and $n$-BP in the suspended, on substrate and encapsulated cases, in units of $10^{-5}$ meVcm$^2$/kV$^2$ / $10^{-7}$ meVcm$^4$/kV$^4$. Results inside (outside) the brackets for $n$-BP are for electric fields applied in $x$ and $y$ directions, respectively.}
{\def\arraystretch{2}\tabcolsep=4pt
\begin{tabular}{c c c c}
\hline\hline
& & Trion & \\
Material & Suspended & On substrate & Encapsulated \\
\hline
MoS$_2$ & 85.323 / 3.191 & 109.317 / 8.698 & 218.286 / 11.227\\
MoSe$_2$ & 86.599 / 4.291 & 107.716 / 9.753 & 204.923 / 12.334\\
WS$_2$ & 122.784 / 3.210 & 204.131 / 7.512 & 376.108 / 7.879\\
WSe$_2$ & 86.809 / 5.742 & 234.332 / 8.575 & 411.367 / 8.480\\
1-BP & 127.649 / 0.531 (14.128 / 0.0088) & 211.017 / 1.513 (23.747 / 0.0637) & 320.562 / 1.827 (36.950 / 0.242)\\
2-BP & 393.922 / 0.436 (68.796 / 0.548) & 508.718 / -0.0239 (90.252 / 1.606) & 621.182 / -1.029 (116.473 / 2.967)\\
3-BP & 639.373 / -2.151 (164.476 / 2.863) & 718.239 / -2.863 (201.404 / 4.778) & 792.271 / -3.699 (250.760 / 6.188)\\
4-BP  & 843.204 / -4.810 (278.942 / 4.475) & 884.925 / -5.218 (330.894 / 5.836) & 928.280 / -5.724 (391.636 / 6.559)\\
\hline\hline
\end{tabular}
}
\label{table:TMDCandBP-Trion}
\end{table*}

Trion binding energies for TMDCs in Table \ref{table:TMDCandBP-TrionBindingEnergy} are all of similar magnitude, namely $\approx$ 30 meV in the suspended case. Moreover, in the absence of electric field, their radii are all $\approx$ 25 \AA\, (see e.g. Fig. \ref{fig:MoS2-Trion}(b), top right panel). With this information, one can estimate the voltage drop at the trion radius, for the highest electric field considered here ($F = $ 60 kV/cm), as $\approx$ 15 meV, which is still around half of the binding energy of the trion. The fact that the voltage drop at the radius of the trion wave function is still smaller than its binding energy explains why the trion is still stable and robust for applied fields up to this value for all TMDC investigated here.

In order to illustrate better the need to fit the trion Stark shift by a fourth (rather than second) order expression, we show the data for suspended MoS$_2$ in Fig. \ref{fig:MoS2-Error} (symbols) along with second and fourth order fitting functions. For the former, we assume two possibilities: a purely quadratic function, i.e. $\Delta E = \frac{1}{2}\alpha F^2$ and a function with an additional linear term (so that the vertex of the parabola is allowed to move away from $F = 0$), $\Delta E = \gamma F + \frac{1}{2}\alpha F^2$. A fitting algorithm, based on damped least-squares method, is used to find the optimal parameters $\gamma$ and $\alpha$ that fit the numerical data. It is clear that the purely quadratic function (dashed) is off the numerical data. The function with $\gamma \neq 0$ provides better fitting, but still not as good as the quartic one, and it is physically less reasonable, since such a linear correction must come from an intrinsic exciton dipole moment in the direction of application of the field (see e.g. Ref. [\onlinecite{Klein}]), which is clearly absent for the ground state excitons discussed here. The best fitting is clearly obtained with the fourth order expression, with a coefficient of determination $R^2 = 0.99996$, which can be compared to the second order expressions with and without the linear term, which have $R^2 = 0.98166$ and $R^2 = 0.95484$, respectively. \cite{Footnote1} The fact that $R^2$ in the fourth order fitting is already very close to 1 (with 10$^-5$ precision) already suggests that higher order corrections are definitely negligible. Similar results are observed for any other material investigated here. 

As for trions in $n$-BP, Stark shifts for the suspended case are shown in Fig. \ref{fig:BP-trion}(a) for a field applied along the $y$ and $x$ directions. Similar to the TMDC cases, the shift requires a non-zero $\beta$ in Eq. (\ref{eq.Stark}) to properly capture the data. Moreover, polarizabilities in this case are direction dependent, as a manifestation of the band structure anisotropy in this material \cite{BPani, Xia}. Hyperpolarizabilities of $n$-BP, as well as those for TMDCs, are all summarized in Table \ref{table:TMDCandBP-Trion}.

Most importantly, trions in $n$-BP are found to be very robust against applied fields, especially in the $y$ (zigzag) direction. The shift observed as $F_y$ reaches 60 kV/cm is negligibly small for suspended monolayer BP (see black squares in Fig. \ref{fig:BP-trion}(a), left panel), and even for higher fields, of the order of 100 kV/cm, the shift still barely reaches $\approx 1$ meV. This is a consequence of the fact that, although the trion binding energy in this case is still of the same order of magnitude as those found in TMDC and $n$-BP, its wave function dramatically spreads along the $x$-direction but negligibly along the $y$-direction, due to mass anisotropy, \cite{ReviewBP, CastellanosGomez, Xia} as one verifies in Fig. \ref{fig:BP-trion}(b), top row of panels. If the field is applied along the $x$-direction, the wave function is clearly deformed, see Fig. \ref{fig:BP-trion}(b), middle row. However, even if a field as high as 100 kV/cm is applied along the $y$-direction, no significant change is observed in the wave function, as one can see by comparing bottom and top rows in Fig. \ref{fig:BP-trion}(b). In fact, the width of the wave function along $y$-direction is $\approx$ 10 \AA\, (see Fig. \ref{fig:BP-trion}(b), top right panel), so that the potential drop at this point with a $F_y$ = 100 kV/cm field is still only $\approx$ 10 meV, much smaller than the trion binding energy. With such a robust trion state, one could use high in-plane electric fields to move the charged exciton accross the material plane without dissociating the excitonic complex, \cite{Footnote} which has implications for possible energy transfer applications.

\section{Conclusions}

We have investigated the effect of an applied in-plane electric field on the binding energies of excitons and trions in TMDCs and few layer BP. Binding energies and polarizabilities are shown to depend on the dielectric environment, but in all investigated cases, excitons and trions binding energies are high enough to allow for their stability in strong fields. Trions in monolayer BP are demonstrated to be especially robust against fields applied along the zigzag ($y$, in Fig. \ref{fig:MoS2-BP-Exciton}(a)) direction, so that even fields as strong as 100 kV/cm produce Stark shifts of only $\approx$ 1 meV and no visible effects on the wavefunction. Moreover, the usual quadratic dependence on the field intensity is demonstrated to be insufficient to describe the Stark shift in all materials investigated here. Fitting results with a fourth order function yields hyperpolarizabilities of the same order of magnitude as those observed, e.g., for excited states in Ba, Ca, and Rb atoms \cite{KulinaBa, KulinaCa, TadaRb}. We believe the results found here will stimulate not only future experimental investigations on the hyperpolarizabilities of trions in 2D materials, but also the development of devices based on transport of charged excitons in 2D materials by applied in-plane electric fields in the near future.

\emph{Node added:} A few months after the submission of this work, a paper came out \cite{Tuan} with a discussion about the accuracy of the Rytova-Keldysh potential used here in the description of trion binding energies. Just like the Rytova-Keldysh potential, the approach in this paper is also based on the solution of Poisson equation for a stack of dielectric slabs, but takes into account the inner structure of the atomic layers of the 2D semiconductor in a more precise way, which is relevant in the context of encapsulated TMDC and provides a correction of a few meV on the trion energies found here for these materials in the absence of electric fields.

\acknowledgments  This work has been financially supported by CNPq, through the PRONEX/FUNCAP and PQ programs. DRR was supported by NSF CHE-1464802.


\begin{references}
\bibitem{Footnote} As a matter of fact, the electric field potential diverges to $-\infty$ as $r \rightarrow \infty$, therefore, any electron-hole pair is expected to eventually dissociate. However, in the cases of strongly bound electrons and holes investisgated here, such dissociation is unlikely to occur before the exciton decays, since it would require tunneling through a thick potential barrier, provided by the strong electron-hole interaction potential.
\end{references}
\end{document}